\documentclass{SciPost}

\hypersetup{
    colorlinks,
    linkcolor={red!50!black},
    citecolor={blue!50!black},
    urlcolor={blue!80!black}
}

\usepackage[bitstream-charter]{mathdesign}
\usepackage{braket}
\DeclareSymbolFont{usualmathcal}{OMS}{cmsy}{m}{n}
\DeclareSymbolFontAlphabet{\mathcal}{usualmathcal}

\fancypagestyle{SPstyle}{
\fancyhf{}
\lhead{\colorbox{scipostblue}{\bf \color{white} ~SciPost Physics }}
\rhead{{\bf \color{scipostdeepblue} ~Submission }}

\fancyfoot[C]{\textbf{\thepage}}
}

\usepackage{xcolor}

\begin{document}

\pagestyle{SPstyle}

\begin{center}{\Large \textbf{\color{scipostdeepblue}{
Replica Tensor Train\\
}}}\end{center}

\begin{center}\textbf{
Miha Srdin\v{s}ek\textsuperscript{1$\star$},
Gabriel Gouraud\textsuperscript{1$\dagger$} and
Xavier Waintal\textsuperscript{1$\ddagger$}
}\end{center}

\begin{center}
{\bf 1} Universit\'e Grenoble Alpes, CEA, Grenoble INP, IRIG, Pheliqs, F-38000 Grenoble, France
\\[\baselineskip]
$\star$ \href{mailto:mihasrdinsek@gmail.com}{\small mihasrdinsek@gmail.com}\,,\quad
$\dagger$ \href{mailto:gab.gouraud@gmail.com}{\small gab.gouraud@gmail.com}\,,\quad
$\ddagger$ \href{mailto:xavier.waintal@cea.fr}{\small xavier.waintal@cea.fr}
\end{center}

\section*{\color{scipostdeepblue}{Abstract}}
\textbf{\boldmath{%
We describe a numerical many-body technique that is based on both tensor networks and quantum Monte Carlo. The variational ansatz is a tensor network that can harvest volume-law entanglement. It is constructed from a tensor train to which one applies a set of non-local operators that force several indices of the tensor train to represent the same physical index, hence its name -- replica tensor train (RTT). From the tensor network toolbox, it inherits the possibility to make linear combinations of these states and apply a certain class of operators. We can therefore find the ground-state of a local Hamiltonian in a purely algebraic way as in standard tensor network algorithms -- i.e. without using gradient descent methods. On the other hand, the volume-law structure forbids calculating physical observables directly. In much the same way as on a quantum computer where one can prepare a state but can only sample it at the end, here we have to use Markov Chain Monte Carlo to compute the observables. We further show that the approach can be extended to build Krylov-subspace ground-state methods within the variational manifold. We illustrate the different algorithms on a two-dimensional spin model with a transverse magnetic field, which can be solved by this approach at low computational cost.
}}

\vspace{\baselineskip}





\section{Introduction}

Finding the ground-state of a quantum many-body problem is hindered by the exponential growth of the Hilbert space size with the number $N$ of particles (e.g., spins, qubits, or fermionic sites). It is very tempting, and often effective, to focus on a small part of the Hilbert space, the one we expect to be relevant, i.e. to use a variational approach to the problem. However, not all variational approaches are born equal. The \emph{computability}, i.e. the mathematical tools at our disposal to compute with them  has strong variations with the underlying class of wave functions. A second thing that  varies is the \emph{expressivity} of the ansatz, i.e. its ability to approximate the true ground-state for a given size of the variational manifold. Often, one must compromise and give up some computability in exchange for an increased expressivity.
 
In terms of computability, the gold standard are arguably the matrix product states (MPS). This one-dimensional tensor network ansatz has everything that one could dream of: it can be sampled exactly, two states can be added together, calculating local observables is straightforward, and we have powerful algebraic approaches to find the ground-state -- i.e., the celebrated DMRG algorithm \cite{White1992a,White1992b}. The only drawback of MPS is that its expressivity is excellent for one-dimensional problems \cite{hastings_area_2007,cirac_matrix_2021}, but somewhat limited for two-dimensional problems and beyond. Its two-dimensional tensor network generalization, PEPS, essentially solves the expressivity problem. However, there is a large cost in computability: finding the ground-state becomes a much more complex affair and the computational cost increases drastically \cite{scarpa_projected_2020}.
 
At the other extreme of the spectrum, one finds ans\"atze with limited internal structure: we only know how to evaluate the unnormalized wave function for a given configuration and, thanks to automatic differentiation, we can easily calculate the gradient of the wave function with respect to its parameters. With such properties, we are left with a unique path for calculating the observables and finding the ground-state: Markov chain Monte Carlo and energy minimization with noisy gradient descent. This approach is known as variational Monte Carlo (VMC) \cite{sorella_green_1998,sorella_green_2000,casula_geminal_2003}. Since there are almost no constraints for constructing a VMC ansatz, highly expressive ones can be built. A natural choice for such constructions is to use a neural network. For instance, convolutional image processing models that retain a notion of space and locality, and language processing transformer models, which describe complex long-range correlations via an attention mechanism. The resulting neural quantum states (NQS) have become very popular \cite{carleo_solving_2017,
nomura_restricted_2017,
vicentini_variational_2019,
choo_two-dimensional_2019,
hibat-allah_recurrent_2020,
lovato_hidden-nucleons_2022,
robledo_moreno_fermionic_2022,
sharir_neural_2022,
kim_neural-network_2023,
pfau_ab_2020,hermann_deep-neural-network_2020,
viteritti_transformer_2023,
chen_empowering_2024,
lange_architectures_2024}. 
Despite the field being rather recent, very promising results have been obtained. Yet, we think that it is fair to state that the noisy gradient descent in a non-convex manifold is not ideal: it requires a large number of optimization steps and there is no guarantee of convergence (presence of local minima or barren plateaus).
 
Researchers on both sides are seeking solutions to this computability-expressivity duality problem. On the NQS side, they try to add more structure. For instance, one can  build objects that automatically capture the antisymmetry of the wave function \cite{robledo_moreno_fermionic_2022} or certain symmetries such as translational invariance \cite{roth_high-accuracy_2023}. On the tensor network side, one can decrease the computational burden by using Markov Chain Monte Carlo on PEPS \cite{wang2011,liu_accurate_2021,liu_tensor_2024,naumann_rizzi_introduction_2024} or increase the expressivity through String Bond States (SBS) \cite{schuch_strings_2008,sfondrini_simulating_2010,glasser_neural-network_2018} or embed MPS inside NQS  \cite{lami_matrix_2022,wang_tensor_2023,wu_tensor-network_2023,srdinsek_hybrid_2026,liang_hybrid_2021,du_neuralized_2025}. This is a very active and quickly evolving field. Indeed, with the techniques already available to us, there is increasing evidence that we are not far from solving the founding models of the field, such as the Hubbard model \cite{qu_phase_2024,xu_coexistence_2024,roth_superconductivity_2025}.
 
 \begin{figure*}[t]
    \centering
    \includegraphics[width = 1.0\linewidth]{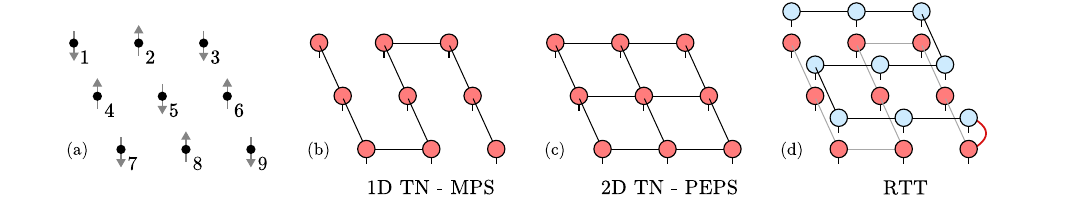}
    \caption{The RTT ansatz compared to other tensor networks (TN). 
    (a) A toy model: a $3\times 3$ grid of $9$ spins. 
    (b) Schematic of an MPS that goes through these spins in a snake-like manner. Some spins may be close by on the lattice but far away in the MPS structure.
    (c) A PEPS ansatz for the same lattice. The connectivity of the PEPS naturally respects the local connections between spins. 
    (d) The RTT ansatz consists of a first (red) MPS followed by a second (blue) MPS such that different sites of the tensor network correspond to a unique physical spin.
    \label{fig:rtt-intro}}
\end{figure*}

The purpose of this article is to address the above problem through a new mathematical object -- we shall call it a Replica Tensor Train (RTT) -- that has interesting mathematical properties. As we shall see, the RTT is an extension of MPS that allows one to quite dramatically increase the expressivity of the MPS while keeping some of its core algebraic properties. The main appeal of RTT, and chief result of this paper, is that while it can harvest volume-law entanglement with low bond dimension, we shall be able to find the ground-state within the RTT manifold \emph{without} resorting to gradient descent. 

In a nutshell, an RTT is a long tensor train where several sites of the tensor network correspond to the same physical spin, see Fig.\ref{fig:rtt-intro} for a schematic. The RTT ansatz is introduced in four different, but equivalent ways in section \ref{sec:rtt}: 
\begin{itemize}
\item as an MPS where physical indices appear more than once,
\item as an ``MPS of MPS'', 
\item as a tensor network, 
\item and as an MPS embedded in a quantum error correction code (An MPS with a set of stabilizers).
\end{itemize}  
We show that these different formulations provide different insights, allowing us to construct our formalism. 

In Section \ref{sec:opt}, we provide the algorithms for finding the ground-state of a given Hamiltonian. The algorithms are rather peculiar: on the one hand, the ground-state can be constructed in a fully algebraic way using only tensor network tools such as tensor contraction and singular value decomposition (as in the DMRG algorithm). On the other hand, to compute any observable, one must sample the state using Markov chain Monte Carlo. In some sense, the situation is very similar to that of a quantum computer where we can construct the state algebraically using quantum gates, but the sampling of qubits is still needed for measuring anything.

In Section \ref{sec:Ising model}, we illustrate the above on a $20\times 20$ two-dimensional transverse-field Ising model.

\section{Replica Tensor Trains}
\label{sec:rtt}

In this section, we build the Replica Tensor Train (RTT). The RTT is a variational ansatz that has special algebraic properties, which will allow us to build new algorithmic procedures for finding ground-states of many-body systems.

\subsection{Preliminary remark: Tensor Trains versus Matrix Product States}

The concepts of Matrix Product States and Tensor Trains are used interchangeably in the literature, the former originating from quantum many-body physics and the latter from mathematics. For the purpose of this article, it will be important to properly separate the two concepts. We define a TT (Tensor Train) $\phi_{i_1,...i_\Omega}$ with $\Omega$ discrete variables $i_1, i_2...i_\Omega$, where $i_\alpha \in \{0,1\}$, (the generalization to dimensions larger than $2$ is trivial) as,
\begin{equation}
\phi_{i_1,...i_\Omega} =
\mathbf{Tr} A_1^{i_1}A_2^{i_2}\dots A_\Omega^{i_\Omega}.
\end{equation}
$A_\alpha^{i_\alpha}$ is a set of two $\chi_\alpha\times\chi_{\alpha+1}$ matrices $A_\alpha^{0}$ and $A_\alpha^{1}$ (or in tensor network language a tensor with three indices) where, by convention, the first and last bond dimensions $\chi_\alpha$ are set to unity: $\chi_1 = \chi_{\Omega+1} = 1$. A TT is just a tensor network. It can represent any large tensor $\phi_{i_1,...i_\Omega}$ without, a priori, any connection to a quantum many-body state.

What we call an MPS, on the other hand, is a TT used to represent a quantum many-body state (this was its actual original meaning before these objects started to be used for many new purposes). We consider a system with $N$ spins, qubits or fermionic occupation states. A basis set for our Hilbert space is $\ket{\boldsymbol{s}}=\ket{s_1...s_N}$ where $s_\alpha \in \{0,1\}$ is the state of the chosen spin/qubit. A general quantum state $\ket{\Psi}$ takes the form,
\begin{equation}
\ket{\Psi} = \sum_{\boldsymbol{s}}\psi_{s_1...s_N} \ket{\boldsymbol{s}}.
\end{equation}
We define the MPS as a state where $\psi_{s_1...s_N}$ is a TT with $\Omega=N$. 

In the rest of this section we will construct other quantum states (the RTT), also defined with TTs, but with $\Omega \ge N$. In exchange for a much higher expressivity (volume law entanglement), these RTT will loose many, but not all, of the nice properties of MPS. In particular, it will not be possible to compute expectations of observables or of the Hamiltonian exactly using contractions; also it will not be possible to rotate the computational basis trivially. However, some crucial properties will remain such as the possiblity to apply an operator or to add two RTT together. 

\subsection{Four equivalent definitions of RTT}

We now construct the Replica TT quantum state. There are four different but equivalent ways of defining the RTT, each providing different insights which will later be useful in the discussion of different algorithms. These definitions are made in terms of a tensor train where physical sites appear more than once (the pedestrian definition, useful to actually evaluate the ansatz), a matrix product state of matrix product state (useful to make contact with previous ansatz), a tensor network (useful to define algeric operations) and a quantum error corrected matrix product state (a curiosity at this stage that could become useful in the future).

\subsubsection{RTT as a TT with repeated appearance of the physical degrees of freedom} 

We define an RTT as follows: we consider a TT $\phi_{i_1...i_\Omega}$ with $\Omega \ge N$. As an example, in the actual applications discussed later, we will have $\Omega = 2N$ or $\Omega = 4N$. We introduce a function $j=\sigma(\omega)$ that takes the index of one the tensor $\omega\in \{1...\Omega\}$ as an input and returns the index of one spin $j\in\{1....N\}$ as an output. The RTT is then defined as,
\begin{equation}
\psi_{s_1...s_N} = 
\mathbf{Tr} A_1^{s_{\sigma(1)}}A_2^{s_{\sigma(2)}}\dots A_\Omega^{s_{\sigma(\Omega)}}.
\end{equation}
In other words, an RTT is a generalization of MPS where each physical degree of freedom may appear more than once, i.e. may be replicated. For later use, we introduce the set $R_j = \sigma^{-1}(j)$ which enumerates the different replica tensors of the TT for a given physical index. Its size is $|R_j|$. The different elements of $R_j$ are designated by $R_{j a}\in\{1..\Omega\}$. By construction $\sum_j |R_j| =\Omega$ and $\sigma(R_{j a}) = j$, $\forall a\in\{1..|R_j|\}$. 

The function $\sigma(\omega)$ is best understood graphically as a discrete "path" between the physical degrees of freedom; the parameter $\omega$ is the "time" of the parametrized curve. Importantly, this path visits the different physical degrees of freedom more than once. Let us illustrate the above construction on a practical example. We consider a small system of $6$ spins on a 2D lattice like so,
\begin{equation}
\vcenter{\hbox{\includegraphics[width=0.4\linewidth]{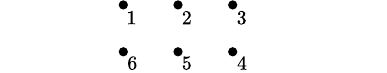}}}\nonumber
\end{equation}
The path $\sigma(\omega)$ with $\Omega=N$ which we call the "horizontal ordering" can be visualized as,
\begin{equation}
\sigma^H(\omega) = \vcenter{\hbox{\includegraphics[width=0.4\linewidth]{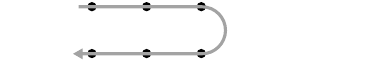}}}
\end{equation}
Since the ordering of the physical sites has been preserved, this path is simply the identity function $\sigma(\omega) = \omega$ and the resulting RTT reduces to a simple MPS,
\begin{equation}
\psi^H_{s_1...s_6} = 
\mathbf{Tr} A_1^{s_1}A_2^{s_2}\dots A_6^{s_6}.
\end{equation}
However, a different ordering of the tensors is also possible. For instance, the "vertical ordering",
\begin{equation}
\sigma^V(\omega) = \vcenter{\hbox{\includegraphics[width=0.4\linewidth]{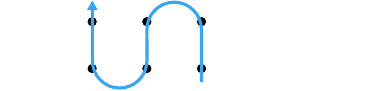}}}
\end{equation}
that also corresponds to a plain MPS but with a different ordering of the tensors,
\begin{equation}
\psi^V_{s_1...s_6} = 
\mathbf{Tr} A_1^{s_4}A_2^{s_3}A_3^{s_2}A_4^{s_5}A_5^{s_6}A_6^{s_1}.
\end{equation}
Our first non-MPS RTT combines both the vertical and the horizontal orderings like so,
\begin{equation}
\sigma^{HV}(\omega) = \vcenter{\hbox{\includegraphics[width=0.4\linewidth]{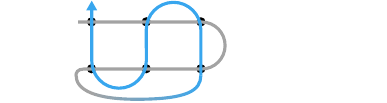}}}
\end{equation}
which corresponds explicitly to the following state:
\begin{eqnarray}
\psi^{HV}_{s_1...s_6} = 
\mathbf{Tr} A_1^{s_1}A_2^{s_2}A_3^{s_3}A_4^{s_4}A_5^{s_5}A_6^{s_6} \nonumber \\
A_7^{s_4}A_8^{s_3}A_9^{s_2}A_{10}^{s_5}A_{11}^{s_6}A_{12}^{s_1}
\end{eqnarray}
We observe in particular that the path goes through all the connections between the spins, i.e. if two spins are neighbors physically, they are also neighbors in \emph{some} part of the RTT. 

For later reference, we define the sum $\sigma = \sigma^1 \oplus \sigma^2$ of two paths $\sigma^1(\omega)$ of size $\Omega_1$ and $\sigma^2(\omega)$ of size $\Omega_2$ as the path of size $\Omega=\Omega_1+\Omega_2$ that goes first through the sites of $\sigma_1$ and then through those of $\sigma_2$:
\begin{equation}
\sigma(\omega) = 
\begin{cases}
\sigma^1(\omega) \ \ \forall \omega\in \{1\cdots\Omega_1\} \\
\sigma^2(\omega-\Omega_1) \ \ \forall \omega\in \{\Omega_1+1\cdots\Omega\}
\end{cases}
\end{equation}
With this notation, we have $\sigma^{HV} = \sigma^{H} \oplus \sigma^{V}$. Another example of path $\sigma$ is shown in Fig.~\ref{fig:rtt-intro}d.

\subsubsection{RTT as a nested MPS of MPS} 

Another way of viewing an RTT is as an "MPS of MPS". For the sake of clarity, let us suppose that the function $\sigma(\omega)$ goes through all the physical sites once, then again through all the sites a second time (in a different order), then again until $\Omega = KN$. We call such a state an RTT with $K$ orderings (with $K$ replicas per site). Defining
\begin{equation}\label{eq:MPS-of-MPS}
\phi_{\boldsymbol{s}}^{(i)} = A_{1+iN}^{s_{\sigma(1+iN)}}A_{2+iN}^{s_{\sigma(2+iN)}}\dots A_{N+iN}^{s_{\sigma(N+iN)}}
\end{equation}
we arrive at,
\begin{equation}
\psi_{\boldsymbol{s}} = \sum_{\alpha_0...\alpha_{K-2}} 
\big[\phi_{\boldsymbol{s}}^{(0)}\big]_{\alpha_0}
\big[\phi_{\boldsymbol{s}}^{(1)}\big]_{\alpha_0\alpha_1}
\cdots
\big[\phi_{\boldsymbol{s}}^{(K-1)}\big]_{\alpha_{K-2}}
\end{equation}
justifying the name ``an MPS of MPS'', as each tensor network $[\phi_{\boldsymbol{s}}^{(i)}]_{\alpha\beta}$ corresponds to a regular MPS. It follows automatically from this remark that RTT generalizes MPS, as well as String Bond States which are a rank 1 product of MPS. However, this construction shows that an RTT does not only allow for a product of different MPS, but also for any linear combinations between them. Last, since it has been shown that String Bond States naturally encompass Restricted Boltzmann Machine \cite{glasser_neural-network_2018}, it follows that RTT contains this class of neural networks as well.

\subsubsection{RTT as a tensor network} 
A third way of viewing an RTT is as a tensor network made of the underlying TT together with Kronecker tensors that enforce the link between the variables of the TT and the physical degrees of freedom.

The Kronecker tensor, also known as the copy tensor, is defined as $\delta_{ijk} = \delta_{ij}\delta_{jk}$. It enforces the fact that all its indices must be equal (and may also be defined with four or more indices). The binding between a spin $s_\alpha$ and the tensors in RTT can be expressed as a contraction between the TT of the RTT and the tensor
\begin{equation}
\delta^{(j)}_{s_j,R_{j}}    = \prod_{a=1}^{|R_j|} \delta_{s_j, R_{j a}}.
\end{equation} 
In this notation, the RTT can be rewritten as the following tensor network contraction,
\begin{equation}
\psi_{\boldsymbol{s}} = \sum_{i_1...i_\Omega}
\mathbf{Tr} A_1^{i_1}A_2^{i_2}\dots A_\Omega^{i_\Omega}
\prod_{j=1}^N \delta^{(j)}_{s_j,R_{j}}
\end{equation} 

Once again, such a contraction is best viewed graphically using standard tensor network drawing conventions. Einstein summation is assumed on any connection between two tensors. For our RTT $\psi^{HV}_{s_1...s_6}$, it reads,
\begin{equation}
\psi^{HV}_{s_1...s_6} = \vcenter{\hbox{\includegraphics[width=0.6\linewidth]{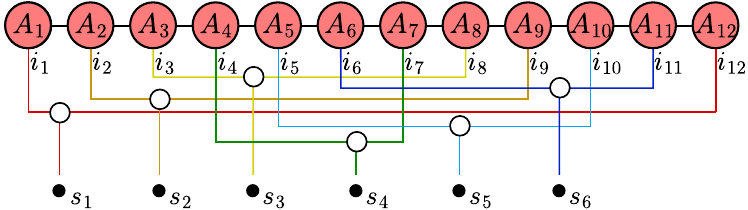}}}\nonumber
\end{equation}
where the empty circles correspond to the Kronecker tensors,
\begin{equation}\label{eq:delta}
\vcenter{\hbox{\includegraphics[width=0.6\linewidth]{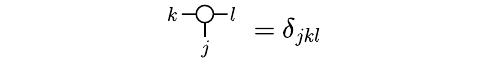}}}.
\end{equation}

To proceed and analyze the structure of this peculiar tensor network, let us name separately the TT part and the Kronecker part. We denote the TT part,
\begin{equation}
\phi_{\boldsymbol{i}} = \phi_{i_1...i_\Omega} = 
\mathbf{Tr} A_1^{i_1}A_2^{i_2}\dots A_\Omega^{i_\Omega}
\end{equation}
where $\boldsymbol{i}$ is a shortcut for $\boldsymbol{i} = (i_1,...i_\Omega)$, and the Kronecker part as,
\begin{equation}
\mathcal{C}_{\boldsymbol{s,i}} = 
\prod_{j=1}^N \delta^{(j)}_{s_j,R_j},
\end{equation}
where $\mathcal{C}$ stands for constraint. With this new notation we have,
\begin{equation}
\label{eq:rtt_equal_C_tt}
\psi_{\boldsymbol{s}} = \sum_{\boldsymbol{i}}\mathcal{C}_{\boldsymbol{s,i}}\phi_{\boldsymbol{i}}
\end{equation}
which can be viewed as a simple matrix-vector multiplication 
\begin{equation}
\psi = \mathcal{C} \phi
\end{equation}
in the exponentially large spaces spanned by $(s_1...s_N)$ and $(i_1...i_\Omega)$ which we call, respectively, the \emph{physical space} and the \emph{replica space}. 

We are now going to make a simple algebraic remark on which are based all the subsequent developments of this article. Suppose that we want to apply an operator $\mathcal{A}$ on our state $\psi$. For instance, $\mathcal{A}$ can be the Hamiltonian $\mathcal{H}$ of the problem but we will also use other operators. Now, further suppose that we know how to commute $\mathcal{A}$ with $\mathcal{C}$. In other words, we suppose that we can compute $\bar{\mathcal{A}}$ such that,
\begin{equation}
\mathcal{A}\mathcal{C} = \mathcal{C}\bar{\mathcal{A}}
\end{equation}
What do we gain from this construction? Essentially, we recover the crucial ability to perform linear combinations of wavefunctions while keeping the ansatz in the same RTT form. Indeed, suppose that we want to construct $\psi' = \psi + \mathcal{A}\psi$. We trivially have $\psi' = \mathcal{C} \phi'$ with $\phi' = \phi + \bar{\mathcal{A}} \phi$. As we shall see below, computing  $\bar{\mathcal{A}} \phi$ will be a straightforward tensor network contraction yielding a new TT; the resulting sum simply amounts to adding two TTs, a standard TT process which yields another TT (see e.g. Section~8.2 of Ref.~\cite{waintal_who_2026}). The capacity to perform these linear combinations is crucial and will provide us with the direct means of finding the ground-state of a problem without resorting to the stochastic gradient descent used in variational Monte Carlo.

We now show that $\bar{\mathcal{A}}$ can be constructed with simple rules when $\mathcal{A}$ is written as a sum of Pauli strings, i.e. is a polynomial in the three Pauli matrices $X_i$, $Y_i$ and $Z_i$ acting on spin $i$. Spin models are naturally written in such a form; fermionic models may be brought to this form using e.g. a Jordan Wigner transformation.  We write $\mathcal{A}$ as,
\begin{equation}
\mathcal{A} = \sum_{k=1}^{N_\mathcal{A}}  \eta_k \mathcal{A}_k
\end{equation}
where $\mathcal{A}_k$ is a Pauli string and $\eta_k$ is the corresponding amplitude. For the special case of strings of two Pauli matrices (the generalization to shorter or longer strings is straightforward),
\begin{equation}
\mathcal{A}_k =  \Lambda^{a_k}_{i_k} \Lambda^{b_k}_{j_k}
\end{equation}
where $\Lambda^a_i$ is the $a^{th}$ Pauli matrix ($a=1: X_i$, $a=2: Y_i$, $a=3: Z_i$) applied on spin $i$. Hence, the integers $a_k$, $b_k$ point to the type of Pauli matrices, and $i_k$, $j_k$ to the spin they apply to. The above decomposition is always possible since Pauli strings form a basis of the space of operators. However, the appeal of this decomposition comes from the fact that for the operators of interest to us, the number of terms $N_\mathcal{A}$ in the decomposition is polynomial (often linear) in $N$.

The rules to compute $\bar{\mathcal{A}}$ follow from the observation that,
\begin{alignat}{3}
X_{jj'} \delta_{j'kl} &= \delta_{jk'l'}X_{kk'}X_{ll'} \nonumber\\
&= \delta_{1-j,kl} \\
Y_{jj'} \delta_{j'kl} &= \delta_{jk'l'}Y_{kk'}X_{ll'}  &=\; &\delta_{jk'l'}X_{kk'}Y_{ll'}\nonumber\\
&= i (-1)^{l}\delta_{1-j,kl}\\
Z_{jj'} \delta_{j'kl} &= \delta_{jk'l'}Z_{kk'}  &=\; &\delta_{jk'l'}Z_{ll'}\nonumber\\
&= (-1)^{l}\delta_{jkl}
\end{alignat}
where we used Einstein summation convention. Again, these rules are best understood graphically:
\begin{align}
\label{eq:commutationx}\vcenter{\hbox{\includegraphics[width=0.4\linewidth]{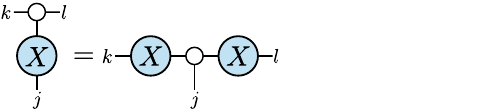}}}\\
\label{eq:commutationy}\vcenter{\hbox{\includegraphics[width=0.4\linewidth]{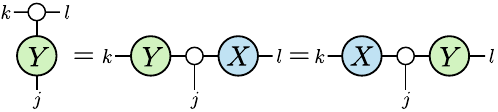}}}\\
\label{eq:commutationz}\vcenter{\hbox{\includegraphics[width=0.4\linewidth]{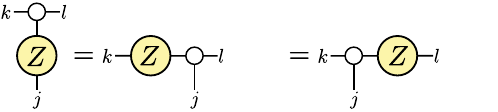}}}
\end{align}
In other words, the rules are the following: when we commute an $X$ operator with $\delta$ we get two $X$ operators (one on each side). On the other hand, when we commute a $Z$ operator with $\delta$, we get a single $Z$ operator on the other side and we have the freedom to choose where we place it. Since $Y = iXZ$, when we commute a $Y$ operator with $\delta$, we get one $Y$ and one $X$ operator on the other side, with the freedom to choose where we place the $Y$. An alternative possibility for the $Z$ operator is to write it as the square ($Z=S^2$) of the "phase gate" $S$ defined as $S = \begin{pmatrix} 1 & 0 \\ 0 & i \end{pmatrix}$. Upon commuting $Z$ with $\delta$, one obtains an $S$ gate on each output leg of the Kronecker tensor.

Let us give a concrete example for our RTT $\psi^{HV}=\mathcal{C}^{HV} \phi^{HV}$. Suppose we want to apply the operator
\begin{equation}
\mathcal{A} = Z_1 Z_2 + Z_1 Z_6 + X_3 + X_1 X_2
\end{equation}
When we commute $Z_1$ with $\mathcal{C}^{HV}$, we can decide to send it to $A_1$ or to $A_{12}$. Same with $Z_6$ which we can send to $A_6$ or $A_{11}$. A possible choice for $\bar{\mathcal{A}}$ is therefore,
\begin{equation}
\bar{\mathcal{A}} = Z_1 Z_2 + Z_{11} Z_{12} + X_3 X_8 + X_1 X_2 X_9 X_{12}
\end{equation}
but alternate choices that satisfy the equation $\mathcal{A}\mathcal{C} = \mathcal{C}\bar{\mathcal{A}}$ are also possible. The above choice has the advantage that it keeps a term local in the physical space also local in the replica space.

\subsubsection{RTT as an MPS with quantum error correction}
 
The fourth way of viewing an RTT is to define it as the projection of an MPS on the logical space of a quantum error correction code \cite{nielsen_quantum_2010}. Indeed, if we define the stabilizers
 \begin{equation}
 V_{j a} = Z_{R_{j ,a}} Z_{R_{j ,a+1}}
 \end{equation}
 then satisfying the constraint $\mathcal{C}$ amounts to enforcing
 \begin{equation}
 \bra{\Psi} V_{j a} \ket{\Psi} = + 1 
 \end{equation}
 for all the stabilizers. In other words, we can write the projector 
 $\mathcal{C}^T\mathcal{C}$ as,
 \begin{equation}
 \mathcal{C}^T\mathcal{C} = \prod_{j=1}^N \prod_{a=1}^{|R_j|-1} \frac{1+ V_{j a}}{2} .
 \end{equation}
In the rest of this article, we will not explore the implications of this stabilizer formulation further. However, it could be an interesting avenue for adapting the algorithms of this article to  actual quantum hardware. Indeed, there exist efficient quantum circuits for constructing any MPS as well as quantum circuits (using ancilla qubits) for measuring the stabilizers.
 
\subsection{Discussion of the expressivity and volume-law entanglement.}

The expressivity of an MPS is directly linked to the value of its bond dimensions since for a given site $a$, the quantum entanglement entropy between the two subsystems defined by $i<a$ and $i\ge a$ is bounded by $S\le \log \chi_a$. The way we order the tensors may also play an important role \cite{moritz_convergence_2004}. More generally, for an arbitrary TT, the bond dimension controls the amount of "correlations" that can be built between two TT sites; any correlation function must decay exponentially with distance with a correlation "length" $\xi$ controlled by $\chi$. In an RTT, the presence of the replicas bring non-local correlations into the picture: they provide "shortcuts" so that two sites that are close in the physical space have (some) replicas that are also close in the replica space. We therefore have two routes to increase the expressivity of the RTT ansatz: increase the bond dimension and/or add new replicas.

We argue that the RTT ansatz can harvest volume-law entanglement. This point is intuitively very clear: by using enough ($\sim N/\xi$) replicas, any physical site can be arbitrarily correlated with any other physical site. The statement can also be understood algorithmically: if one tries to compress an RTT back into an MPS form using standard tools (writing e.g. the constraints as rank 2 matrix product operators), one expects an exponential increase of the bond dimension. A more formal proof of the assertion follows from previous results of the literature. Indeed, seeing the RTT as a nested MPS of MPS [see Eq.~\eqref{eq:MPS-of-MPS}] it follows that "string bond states" are a particular case of RTT. Since these string bond states are themselves a generalization of the Restricted Boltzmann Machine (RBM) \cite{glasser_neural-network_2018} and the RBM can exhibit volume-law entanglement \cite{deng2017}, it follows that RTT possesses this characteristic too. In that point of view, the replicas play the role of the hidden units in an RBM.

In this article, we restrict ourselves to very specific paths [shown in Eqs.~(\ref{eq:orderings},~\ref{eq:orderings_vhddp})], however, many other paths can be considered depending on the entanglement structure of the targeted quantum state. In the following, we explore how the RTT representation of volume-law states can be learned using different algebraic methods, avoiding the gradient descent typically used in volume-law ans\"atze such as RBMs and NQS.
 
\section{Finding the ground-state with an RTT ansatz}
\label{sec:opt}

In this section, we discuss our approach to find the ground-state $\psi_0$ of a Hamiltonian $\mathcal{H}$ using an RTT variational ansatz. As we shall see, the technique is very peculiar: the ground-state can be constructed in a purely algebraic way without resorting to gradient descent (in contrast to what is done in VMC). In that sense, the technique to find the ground-state is relatively close to DMRG. On the other hand, in order to compute observables, we need to resort to sampling, i.e. the method becomes stochastic. In that sense, the technique shares similarities with what is usually done in VMC. 
 
Perhaps a good analogy is with quantum computers where the construction of the state is purely deterministic through the application of a sequence of quantum gates. But in order to measure anything, one needs to measure the qubits and the output becomes stochastic.

We will design two distinct routes for obtaining the ground-state. In the first, we work purely in the replica space while in the second we work in the physical space. What we mean by this statement is that in the first technique, we work directly on the object $\phi$ using the replica space Hamiltonian $\bar{\mathcal{H}}$. In contrast, in the second technique, we work on $\psi$ using the actual physical Hamiltonian $\mathcal{H}$.

\subsection{Optimization in the replica space}
\label{sec:hidden-optimization}

Our first approach is to use a variant of the power method to find the ground-state of $\mathcal{H}$. Starting from an initial RTT $\psi^{(0)} = \mathcal{C} \phi^{(0)}$, we define
\begin{equation}
\psi^{(P)} =  (1-\tau \mathcal{H})^P \psi^{(0)}
\end{equation}
By construction (provided $\tau$ is sufficiently small and the overlap of $\psi^{(0)}$ with $\psi_0$ is nonzero), $\psi^{(P)}$ converges to the ground-state $\psi_0$,
\begin{equation}
\psi_0 = \lim_{P\rightarrow\infty} \psi^{(P)} 
\end{equation}
We seek to represent $\psi_0$ as an RTT, by using,
\begin{equation}
\psi^{(P)} =
\mathcal{C} \phi^{(P)}
\end{equation}
with,
\begin{equation}
 \phi^{(P)} \equiv
 (1-\tau \bar{\mathcal{H}})^P \phi^{(0)}
\end{equation}
So far, the process is exact. However, in its naive implementation, the internal rank $\chi$ of $\phi^{(P)}$ may increase very quickly with $P$. The actual algorithm alternates the application of $1-\tau \bar{\mathcal{H}}$ with compression steps. Indeed, the underlying assumption of the overall approach is that $\psi_0 $ can be approached very precisely with a small value of $\chi$.

More precisely, $\phi^{(P)}$ is created from $\phi^{(P-1)}$ using the following sequence,
\begin{itemize}
\item compute the different contributions $\eta_i \bar{\mathcal{H}}_i \phi^{(P-1)}$ for $\forall i \in \{1...N_{\bar{\mathcal{H}}} \}$. The Pauli strings are MPOs with rank 1 so these MPO.MPS multiplications are straightforward and do not increase the rank.
\item sum the different terms together two by two.
\item compress the resulting sums to a targeted rank $\chi$ (see the algorithm in e.g. Section~5.3.1 of Ref.~\cite{waintal_who_2026}).
\item repeat the two preceding steps until all the terms are exhausted and one is left with a single TT.  Multiply the result by $\tau$ and subtract it from $\phi^{(P-1)}$ to obtain $\phi^{(P)}$.
\end{itemize}

In other words, the algorithm amounts to performing a form of Time-evolving block decimation (TEBD) with the replica space Hamiltonian $\bar{\mathcal{H}}$. Other approaches are also possible. For instance, one could try and run the DMRG algorithm on $\bar{\mathcal{H}}$ using an auto-MPO computation followed by a regular DMRG computation leveraging the fact that if $\phi$ is an eigenstate of $\bar{\mathcal{H}}$ then $\psi = \mathcal{C} \phi$ is an eigenstate of $\mathcal{H}$. In our preliminary exploration, we have found that our TEBD-like approach is effective and although it might be improved, the crucial point does not lie there but in the choice of $\bar{\mathcal{H}}$. Indeed, as we have mentioned above, there are many different choices of $\bar{\mathcal{H}}$ that correspond to a single $\mathcal{H}$. While all are valid choices, the different TT $\phi$ are not all amenable to the same level of compression. We will experiment with different possible choices in the next section.

\subsection{Optimization in the physical space}
\label{sec:physical-optimization}

As we shall see, the preceding technique has some limitations in terms of precision in its current form. The reason is that the algorithm works blindly: one does not compute the energy of the state until the very end, so the effect of the compression on the energy is unknown and cannot be mitigated.

While the algorithm in the preceding subsection was close in spirit to the power method, we now introduce an algorithm that resembles the Lanczos Krylov technique. Let $\psi$ be our current approximation of the ground-state $\psi_0$ and let $\mathcal{O}_\alpha$ be a set of operators. For instance, to recover a variant of the Lanczos algorithm, we use $\mathcal{O}_\alpha = \mathcal{H}^{\alpha-1}$ but we will consider also other operators. By convention $\mathcal{O}_1 = \mathbb{1}$ is the identity operator. The algorithm goes as follows. First, we construct the products $\psi^\alpha = \mathcal{O}_\alpha \psi$, by using $\bar{\mathcal{O}}_\alpha$, as done in the replica space optimization. Second, we seek the new approximation of the ground-state $\psi'$ in the form of an optimal linear combination of these states,
\begin{equation}
\label{eq:lin_comb}
\psi' = \sum_\alpha c_\alpha \psi^\alpha.
\end{equation}
Because RTTs are by construction \emph{not} normalized, we also need to introduce the constraint that the state has to be normalized. $\psi'$ is then the solution of the generalized eigenvalue problem 
\begin{equation}
\label{eq:gen_eig1}
H c = \lambda S c,
\end{equation} 
with the lowest eigenvalue. The Hamiltonian matrix $H_{\alpha\beta}$ and the overlap matrix $S_{\alpha\beta}$ are expressed in the reduced basis
\begin{eqnarray}
H_{\alpha\beta} &=& \psi^{\alpha\dagger} \mathcal{H} \psi^\beta \\
S_{\alpha\beta} &=& \psi^{\alpha\dagger}  \psi^\beta,
\end{eqnarray}
with the normalization enforced by $c^\dagger S c = 1$, where $c= (c_1 c_2...)$ represents the vector of the coefficients $c_\alpha$. This small eigenvalue problem (we typically consider fewer than ten operators $\mathcal{O}_\alpha$) is readily solved and it yields the new approximation $\psi'$ of the ground-state. If we wish to continue with another optimization step, we turn $\psi'$ back to an RTT form by summing the TTs $\phi_\alpha$. The final $\phi'$ can also be compressed.

The important difference between this scheme and the one of the preceding section is that we need to compute the matrix elements $H_{\alpha\beta}$ and $S_{\alpha\beta}$. These computations will involve a stochastic approach. Indeed, while in principle these matrix elements could be computed through a direct contraction of the associated tensor network, it is unfeasible in practice as the cost is exponential in $N$. In the next section, we show how a simple adaptation of the variational Monte Carlo algorithm, which we call dual Monte Carlo (because it involves sampling both $|\psi_\alpha|^2$ \emph{and} $|\psi_\beta|^2$), solves this problem. 

\subsection{Dual Monte Carlo}
\label{sec:monte-carlo}

The remaining problem is the following: given two RTTs $\psi^1$ and $\psi^2$, how can we compute the matrix elements $\psi^{1\dagger} \mathcal{H} \psi^2$ and $\psi^{1\dagger} \psi^2$. The following algorithm is actually very general and does not require the states to be RTTs, it would work equally well for e.g. neural quantum states. Although not widely spread, similar algorithms have already been developed in the literature \cite{nightingale_optimization_2001,umrigar_alleviation_2007}.

Let us call $\mathcal{Z}_1 = \psi^{1\dagger} \psi^1 = \sum_{\boldsymbol{s}} \psi^{1*}_{\boldsymbol{s}} \psi^1_{\boldsymbol{s}}$ and $\mathcal{Z}_2 = \psi^{2\dagger} \psi^2$ the unknown norms of the corresponding states. Using Markov Chain Monte Carlo (e.g. the Metropolis algorithm), we can sample the probability distribution $P_{\boldsymbol{s}}$ defined as,
\begin{equation}
P^1_{\boldsymbol{s}} = \frac{|\psi^1_{\boldsymbol{s}}|^2}{\mathcal{Z}_{1}}
\end{equation}
with a similar definition for $P^2_{\boldsymbol{s}}$. Hence, we can obtain the desired quantities by sampling over $P^1$ and $P^2$ but only up to the unknown normalization factors $\mathcal{Z}_{1}$ and $\mathcal{Z}_{2}$. To see how we can get rid of these normalizations, let us first look at the matrix elements of the normalized states
\begin{eqnarray}
\label{eq:dual1}
\frac{\psi^{1\dagger}}{\sqrt{\mathcal{Z}_1}} \mathcal{H} \frac{\psi^2}{\sqrt{\mathcal{Z}_2}} &=& 
\frac{1}{\sqrt{\mathcal{Z}_1\mathcal{Z}_2}}
\sum_{\bar{\boldsymbol{s}},\boldsymbol{s}}
\psi^{1*}_{\boldsymbol{s}} \mathcal{H}_{\boldsymbol{s},\bar{\boldsymbol{s}}} 
\psi^2_{\bar{\boldsymbol{s}}}
\nonumber\\
&=&\sqrt{\frac{\mathcal{Z}_1}{\mathcal{Z}_2}} \left\langle \sum_{\bar{\boldsymbol{s}}}
\frac{1}{\psi^1_{\boldsymbol{s}}}
\mathcal{H}_{\boldsymbol{s},\bar{\boldsymbol{s}}} 
\psi^2_{\bar{\boldsymbol{s}}}
\right\rangle_{P^1_{\boldsymbol{s}}}.
\end{eqnarray} 
where readers familiar with VMC will recognize inside the average $\langle\cdots\rangle_{P_{\boldsymbol{s}}^1}$ over the distribution $P^1$ a quantity similar to the local energy used in VMC. We shall call it a \emph{mixed local energy} as both wavefunctions are involved in its calculation (in contrast to the standard VMC definition). By multiplying the above equation with the same equation where the roles of $\psi^1$ and $\psi^2$ are reversed, we see that the normalization factor cancels out
\begin{equation}
\left|\frac{\psi^{1\dagger}}{\sqrt{\mathcal{Z}_1}} \mathcal{H} \frac{\psi^2}{\sqrt{\mathcal{Z}_2}}\right|^2 = 
\bigg\langle \sum_{\bar{\boldsymbol{s}}}
\frac{1}{\psi^1_{\boldsymbol{s}}}
\mathcal{H}_{\boldsymbol{s},\bar{\boldsymbol{s}}} 
\psi^2_{\bar{\boldsymbol{s}}}
\bigg\rangle_{P^1_{\boldsymbol{s}}}
\bigg\langle \sum_{\bar{\boldsymbol{s}}}
\frac{1}{\psi^2_{\boldsymbol{s}}}
\mathcal{H}_{\boldsymbol{s},\bar{\boldsymbol{s}}} 
\psi^1_{\bar{\boldsymbol{s}}}
\bigg\rangle_{P^2_{\boldsymbol{s}}}
\end{equation}
In other words, we need to perform \emph{two} Monte Carlo simulations to obtain the matrix element: one sampling $P^1$ and one sampling $P^2$. The phase of the matrix element is recovered from the phase of Eq.\eqref{eq:dual1} since $\sqrt{\mathcal{Z}_1/\mathcal{Z}_2}$ is a real positive number. The overlaps can be analogously calculated by replacing $\mathcal{H}_{\boldsymbol{s},\bar{\boldsymbol{s}}}$ with $\delta_{\boldsymbol{s},\bar{\boldsymbol{s}}}$ in the above equation yielding,
\begin{equation}
\left|\frac{\psi^{1\dagger}}{\sqrt{\mathcal{Z}_1}}\frac{\psi^2}{\sqrt{\mathcal{Z}_2}}\right|^2 = 
\left\langle 
\frac{\psi^1_{{\boldsymbol{s}}}}{\psi^2_{\boldsymbol{s}}}
\right\rangle_{P^2_{\boldsymbol{s}}}
\left\langle 
\frac{\psi^{2}_{{\boldsymbol{s}}} }{\psi^1_{\boldsymbol{s}}}\right\rangle_{P^1_{\boldsymbol{s}}}.
\end{equation} 

\begin{figure}[t]
    \centering
    \includegraphics[width = 0.7\linewidth]{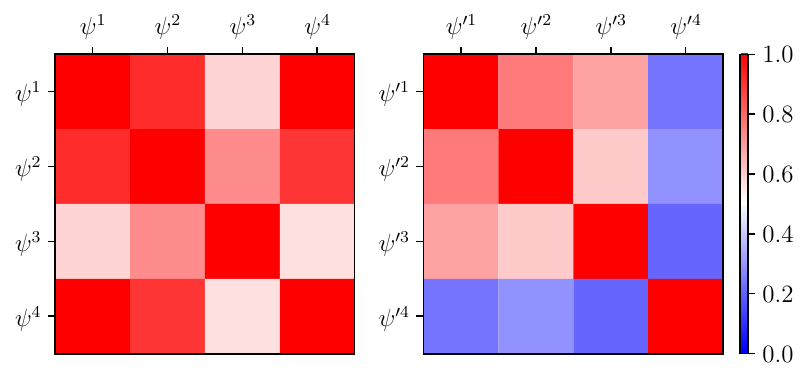 }
    \caption{Example of the $S$ matrix computed with dual Monte Carlo before (left) and after (right) the first step of the orthogonalization procedure. The matrix was constructed when solving the Ising model in Fig.~\ref{fig:dmc-opti}. The new states are compressed to $\chi=2$. \label{fig:orthogonalization}}
\end{figure}
We note that there are two situations where the above expression may lead to vanishing overlap: if the two wavefunctions do not have common support or if they have completely unrelated phases. In thess two cases, the off-diagonal matrix element $S_{\alpha\beta}$ is very small which is not an issue in itself since the diagonal part $S_{\alpha\alpha}$ is equal to one. However, this vanishing overlap will likely be acompanied by vanishing matrix elements $H_{\alpha\beta}$ so that the Krylov problem will bring no gain. This is not an issue of the dual Mont-Carlo algorithm but of the choise of operators $\mathcal{O}_\alpha$. We have not encountered these situations so far in our numerical experiments.

In our formulation of the generalized eigenvalue problem, we shall not work with the normalized states above, but directly with the unnormalized ones. To this end we define
\begin{eqnarray}
\tilde{H}_{\alpha\beta} &=& \left\langle \sum_{\bar{\boldsymbol{s}}}
\frac{1}{\psi^\alpha_{\boldsymbol{s}}}
\mathcal{H}_{\boldsymbol{s},\bar{\boldsymbol{s}}} 
\psi^\beta_{\bar{\boldsymbol{s}}}
\right\rangle_{P^\alpha_{\boldsymbol{s}}} \\
\tilde{S}_{\alpha\beta} &=& \left\langle 
\frac{\psi^{\beta}_{{\boldsymbol{s}}}}{\psi^{\alpha}_{\boldsymbol{s}}}
\right\rangle_{P^\alpha_{\boldsymbol{s}}}
\end{eqnarray}
which implies that we have ${H}_{\alpha\beta} = \mathcal{Z}_\alpha \tilde{H}_{\alpha\beta}  = \tilde{H}^\dagger_{\alpha\beta} \mathcal{Z}_\beta$ and ${S}_{\alpha\beta} = \mathcal{Z}_\alpha \tilde{S}_{\alpha\beta} = \tilde{S}^\dagger_{\alpha\beta} \mathcal{Z}_\beta$. Dividing each row $\alpha$ of Eq.\eqref{eq:gen_eig1} by $\mathcal{Z}_\alpha$, we arrive at a new generalized eigenvalue problem
\begin{equation}
\label{eq:gen_eig2}
\tilde{H} c = \lambda \tilde{S} c,
\end{equation}
expressed in terms of the un-symmetrized Hamiltonian matrix $\tilde{H}$ and the un-symmetrized overlap matrix $\tilde{S}$. The result of this problem is normalized as $\sum_{\alpha\beta} c_\alpha^* \tilde{S}_{\alpha\beta} c_\beta = 1$, so the resulting state $\psi'$ is not normalized. However, one can verify that the normalizations cancel in the calculation of the energy so that the lowest generalized eigenvalue $\lambda_0$ remains the actual energy of the system.
\begin{equation}
E = \frac{\langle \Psi'| \mathcal{H} | \Psi'\rangle}
{\langle \Psi'| \Psi'\rangle} = \lambda_0
\end{equation}
In practice, we need to consider a slight modification of the above for numerical stability. Indeed, as one proceeds to find the ground-state, all the states $\psi^\alpha$ will approach $\psi_0$, so that the overlap matrix $S_{\alpha\beta}$ becomes almost singular: $S_{\alpha\beta} \approx 1 \ \ \forall \alpha,\beta$ making the generalized problem unstable. To solve this problem, we perform a stochastic generalization of the Gram-Schmidt orthogonalization of the states $\psi_\alpha$ with respect to $\psi_\beta$  $\forall \beta <\alpha$, so that we obtain an approximately diagonal matrix $S_{\alpha\beta}$. We define the new states $\psi^{\prime\alpha}$ iteratively, as
\begin{align}\label{eq:Gram-Schmid}
\psi^{\prime 1} &= \psi^{1} \\
\psi^{\prime 2} &= \psi^{2} - \left\langle\frac{\psi^{2}}{\psi^{1}}\right\rangle_{P^{1}}\psi^{1} \\
&\vdots \nonumber\\
\psi^{\prime M} &= \psi^{M} - \sum_{\alpha=1}^{M-1}\left\langle\frac{\psi^{M}}{\psi^{\alpha}}\right\rangle_{P^{\alpha}}\psi^{\alpha},
\end{align}
where $M$ is the number of states $\psi^{\alpha}$. In practice, we have found that it is often sufficient to re-orthogonalize with respect to the first vector $\psi^{1}$ only. As an example of a dual Monte Carlo calculation, a matrix $S_{\alpha\beta}$ before (left) and after (right) re-orthogonalization is shown in Fig.\ref{fig:orthogonalization}.

This concludes the general description of the algorithm. We note that although dual Monte Carlo is very handy, it is not strictly speaking necessary: since the optimization performed on the $c_\alpha$ is done on a handful of parameters, one could use a standard variational Monte Carlo for this very small problem (a typical number we have used so far is $M\le 5$). The key aspect of the physical optimization is our ability to select a handful of important $M$ directions for finding the ground-state (typically guided by the Hamiltonian), as opposed to using a gradient on thousands or hundreds of thousands of parameters.

\section{Application to the 2D transverse Ising model}
\label{sec:Ising model}

We now turn to an actual calculation to illustrate the above RTT approach on a practical example. We consider the transverse field Ising model with nearest-neighbor interactions  on a square lattice of $N = L\times L$ spins. The Hamiltonian is defined as, 
\begin{equation}\label{eq:ising-model}
\mathcal{H} =\sum_{\langle i,j\rangle}Z_{i}Z_{j}-h_{x}\sum_{i}X_{i},
\end{equation} 
where $\langle i,j\rangle$ represents the sum over all the nearest-neighbor pairs of sites. We use open boundary conditions. The interest in this model stems from different aspects. First, as a function of the transverse magnetic field $h_{x}$ the model undergoes a quantum phase transition, from the paramagnetic to the antiferromagnetic phase which breaks the $\mathbb{Z}_{2}$ symmetry \cite{webber_rayleigh-gauss-newton_2022,vecsei_lee-yang_2022}. This is a paradigmatic example of a quantum phase transition. The transition is furthermore accompanied by a classical phase transition at finite temperature. Second, these types of models have recently received a lot of interest in the context of quantum annealing \cite{scholl_quantum_2021,ebadi_quantum_2021,chen_continuous_2023}. The quantum annealers based on Rydberg atoms, in particular, are described by Eq.~\eqref{eq:ising-model} with slightly longer range interactions. Third, we have already established in a previous publication that String Bond States with very low rank are extremely efficient ans\"atze for this model \cite{srdinsek_hybrid_2026}. Since RTT generalizes String Bond States, it follows that RTT must as well be very efficient for this problem. Our goal below is merely to illustrate the new algebraic approach described in the preceding section, we postpone the study of more difficult models to future publications.  

\subsection{Specification of the algorithm}
To specify the algorithm entirely, we need to specify two things: the path (i.e. the function $\sigma(\omega)$) and the operators $\mathcal{O}_\alpha$.

\subsubsection{Specification of the path}
We use a path that first goes through all the sites column by column ($V$ for Vertical), then line by line ($H$ for Horizontal), then diagonally ($D$) then along the other diagonal ($D'$).

More formally, we introduce the different paths  $\sigma^V$, $\sigma^H$, $\sigma^D$ and  $\sigma^{D'}$ defined graphically as,
\begin{equation}\label{eq:orderings}
\vcenter{\hbox{\includegraphics[width=0.6\linewidth]{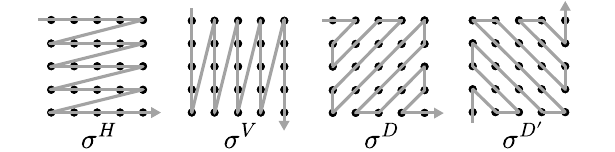}}}.
\end{equation}
The global path $\sigma(\omega)$ of  our RTT is
\begin{equation}
\label{eq:orderings_vhddp}
\sigma = \sigma^{HVDD'} \equiv \sigma^V \oplus \sigma^H \oplus \sigma^D \oplus \sigma^{D'}
\end{equation}
It follows that for a given site $i$, $R_{i1}$ is the TT index of this site in the horizontal ordering, $R_{i2}$ corresponds to $V$, $R_{i3}$ to $D$ and $R_{i4}$ to $D'$.
The open paths that we used are well adapted to the open boundary conditions of the model. These paths would require adaptation to be used for a model with periodic boundary conditions in order to keep the prescription that "sites close by physically should have replicas that are close by somewhere in the replica space".

\subsubsection{Definition of the operators in the physical space}
Second, we need to define the operators $\mathcal{O}_\alpha$. We use only local operators that include at most two Pauli matrices. Our convention is the following: we write $\alpha = (A,B)$ where $A$ indicates which Pauli matrix is concerned (for instance $A=ZZ$ for product of two $Z$ operators and $A=X$ for a single $X$ operator). For pairs of Pauli strings, $B$ labels the direction of the bond ($V$, $H$, $D$, $D'$, possibly combinations such as $VH$ if both vertical and horizontal bonds are considered). For instance, nearest neighbour $ZZ$ operators along the vertical direction are noted,
\begin{equation}
\mathcal{O}_{ZZ, V} = \sum_{i = (i_x,i_y)}  Z_{(i_x,i_y)} Z_{(i_x,i_y+1)} 
\end{equation}
while along the diagonal direction, they are noted,
\begin{equation}
\mathcal{O}_{ZZ, D} = \sum_{i = (i_x,i_y)}  Z_{(i_x,i_y)} Z_{(i_x+1,i_y+1)}.
\end{equation}
Likewise the transverse field operator reads,
\begin{equation}
\mathcal{O}_{X} = \sum_{i = (i_x,i_y)}  X_{(i_x,i_y)}  
\end{equation}
With these notations, the transverse field Ising Hamiltonian reads,
\begin{equation}
\mathcal{H} = \mathcal{O}_{ZZ, V} + \mathcal{O}_{ZZ, H}
-h_x \mathcal{O}_{X}.
\end{equation}

\subsubsection{Definition of the operators in the replica space}
Next, when we commute the $Z_i$ operators with the Kronecker circuit $\mathcal{C}$, we have several choices for the definition of $\bar{\mathcal{O}}_\alpha$ i.e. we can decide where we move the $Z_i$ operators. We use the notation $\rightarrow C$ where $C = V,H,D,D'$ to indicate on which part of the RTT the $Z_i$ are sent. This is not the most general choice, here we assume that all the $Z_i$ operators of a given operator are sent to the same part of the RTT. For instance, for the example of the $\mathcal{O}_{ZZ, V}$ operator, we might decide to send the $Z_i$ to the vertical part of the path (this is the natural path that works best in practice),
\begin{equation}
\bar{\mathcal{O}}_{ZZ, V\rightarrow V} = \sum_{i = (i_x,i_y)}  Z_{R_{i,2}} Z_{R_{i=(i_x,i_y+1),2}} 
\end{equation}
but we could also decide to send them to the diagonal part of the path,
\begin{equation}
\bar{\mathcal{O}}_{ZZ, V\rightarrow D} = \sum_{i = (i_x,i_y)}  Z_{R_{i,3}} Z_{R_{i=(i_x,i_y+1),3}} 
\end{equation}
There is no such freedom for the $\mathcal{O}_{X}$ operator, so we simply have,
\begin{equation}
\bar{\mathcal{O}}_{X} = \sum_{i = (i_x,i_y)}  
X_{R_{i,1}}  X_{R_{i,2}}X_{R_{i,3}}X_{R_{i,4}}.
\end{equation}
Note that in these numerical experiments we limit ourselves to using operators in the replica space that are obtained using the commutation rules derived earlier. This is important for replica space optimization to guarantee that the replica Hamiltonian has the same eigenstates as the physical one. However, no such constraint exists for physical space optimization where we may use any operator we want. This freedom may be used for e.g. optimizing a particular tensor $A^{(i)}$ (or a pair of tensors) without resorting to gradient descent. Another remark is that if all the operators contain only $Z_i$ Pauli matrices, then Dual Monte Carlo is not needed -- a single Monte Carlo simulation provides all the needed matrix elements (since all the different states share the same probabilities).

\subsection{Replica space optimization}   
\begin{figure}[t]
    \centering
    \includegraphics[width = 0.7\linewidth]{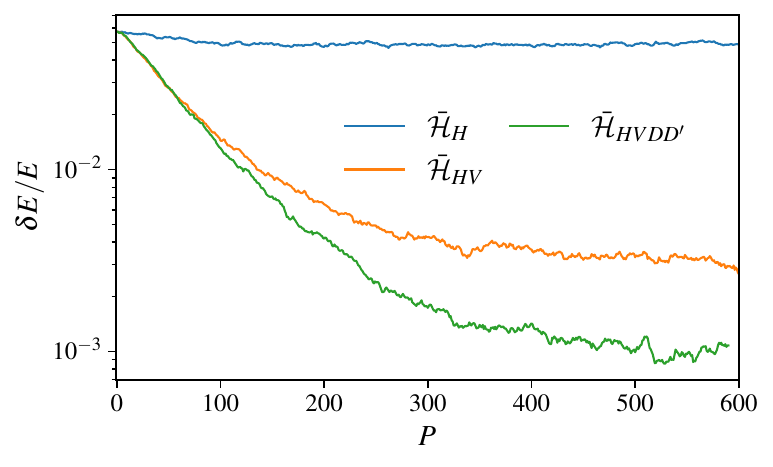}
    \caption{Error on the energy during a replica space optimization by application of $(1-\tau \bar{\mathcal{H}})^P$ for three choices of $\bar{\mathcal{H}}$, see text. The number of active sub-MPS is one (blue), two (orange) and four (green). The optimization is purely algebraic but the calculation of the energy is performed using VMC sampling. The parameters are $h_{x}=3$, $L=20$, $\chi=2$, $\tau=0.1$ and $\alpha = 0.08$ for the $\bar{\mathcal{H}}_{HVDD'}$ curve (green line). \label{fig:rtt-descent}}
\end{figure}

\begin{figure}[t]
    \centering
    \includegraphics[width = 0.7\linewidth]{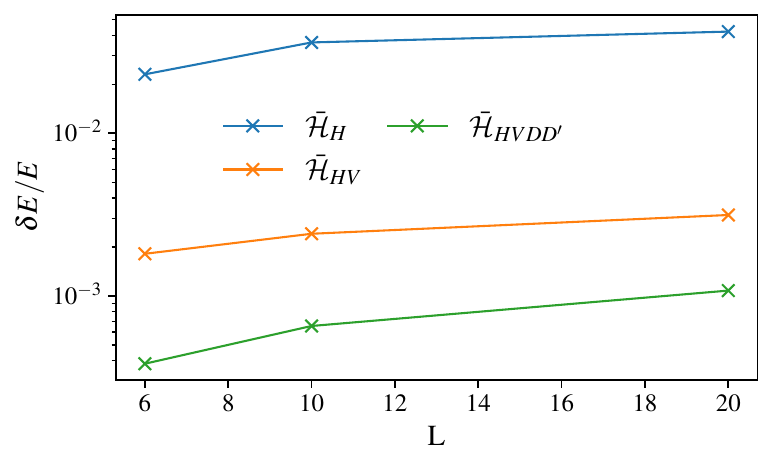}
    \caption{Error on the energy after a replica space optimization versus system size $L$. The number of active sub-MPS is one (blue), two (orange) and four (green), see text. The accuracy of the RTT is robust with respect to system size. Same parameters as in Fig.\ref{fig:rtt-descent}.  \label{fig:large_L}}
\end{figure}

We now present the result of some numerical experiments where we apply the operator $1-\tau \bar{\mathcal{H}}$ iteratively $P$ times on an initially constant TT $\phi^{(0)}$ ($\phi^{(0)}_\mathbf{i} = 1\ \ \forall \mathbf{i}$). Importantly, the algorithm to obtain $\phi^{(P)}$ is perfectly deterministic. However, to obtain the physical observable (here the energy), we need to resort to VMC stochastic sampling. In all the remaining, the corresponding statistical errors are always negligible compared to the error due to the ansatz being imperfect. Only the latter is discussed. The results are shown in Fig.~\ref{fig:rtt-descent}  for $h_{x}=3$, a point very close to the quantum critical point \cite{webber_rayleigh-gauss-newton_2022,vecsei_lee-yang_2022}, hence supposedly the most difficult to compute. The reference ground-state energy $E/N=-3.18185\pm0.00004$ for this point was obtained using the Green Function Monte Carlo algorithm and a perceptrain ansatz as discussed in \cite{srdinsek_hybrid_2026}. We choose the smallest non-trivial rank for the RTT, $\chi=2$. One should keep in mind that tensor network calculations are typically performed with much higher ranks in the hundreds or thousands. We use three different choices of $\bar{\mathcal{H}}$. The first is 
\begin{equation}
\label{eq:H_H}
\bar{\mathcal{H}}_{H} = \bar{\mathcal{O}}_{ZZ,H\to H} + \bar{\mathcal{O}}_{ZZ,V\to H}-h_{x}\bar{\mathcal{O}}_{X}\\
\end{equation}
This choice corresponds to sending all the operators to a single sub-MPS part of the RTT. Therefore, it does not exploit the full freedom of the RTT but rather works as a regular MPS. We find that the error barely decreases to an error $\delta E/E \approx 6\times 10^{-2}$ consistent with what one obtains with a regular MPS for such a small rank. The situation improves drastically when one sends the different interaction terms to the parts of the RTT where they are local, i.e. when we use,
\begin{equation}
\label{eq:H_HV}
\bar{\mathcal{H}}_{HV} = \bar{\mathcal{O}}_{ZZ,H\to H} + \bar{\mathcal{O}}_{ZZ,V\to V}-h_{x}\bar{\mathcal{O}}_{X}.
\end{equation}
We find that the error drops by more than one order of magnitude reaching $\delta E/E \approx 3\times 10^{-3}$, an error level where the physics of the ansatz is qualitatively correct. It is still possible to improve on these results in a variational manner in order to take advantage of the diagonal parts of the RTT $D$ and $D'$. For instance, if we add one more operator with  a parameter $\alpha$,
\begin{equation}
\bar{\mathcal{H}}_{HVDD'} = \bar{\mathcal{H}}_{HV}
+ \alpha \left(
\bar{\mathcal{O}}_{ZZ,D\to D} + \bar{\mathcal{O}}_{ZZ,D'\to D'} \right),
\end{equation}
we find that the error reaches $\delta E/E \approx 10^{-3}$ for the optimum choice of the parameter $\alpha = 0.08$. In the various experiments that we have performed, we have found that the curves shown in Fig.~\ref{fig:rtt-descent} are very generic: the descent is fast, robust and independent of the initial state or choice of metaparameters. The error levels that we have achieved are consistent with those found using VMC and gradient descent using the technique described in \cite{srdinsek_hybrid_2026} for, respectively, $K=1$ (MPS level), $K=2$ (two subparts of the RTT are active) and $K=4$ (the diagonal parts are also active). The advantage of the present approach is that we bypass entirely the non-convex optimization problem. We have also checked that the method works robustly across different system sizes in Fig.~\ref{fig:large_L} where we observe that the error barely increases while going from a $6\times 6$ system to a $20\times 20$ one. This is in sharp contrast to plain MPS where the error deteriorates dramatically at fixed rank (see section \ref{sec:other}). For the $\bar{\mathcal{H}}_{HVDD'}$ replica space Hamiltonian, the value $\alpha=0.08$ was found to be close to optimum across the three system sizes. The optimum value of $\alpha$ was found by a simple 1D scan of $\alpha$, these calculations being rather light computationally. 

\subsection{Physical space optimization}
\begin{table}[t]
\centering
\caption{\label{tab:different-operators}
Energy and energy variance after the physical space optimization. The $\times$ indicate which operators have been included. The starting point is the replica space optimization using the $\bar{\mathcal{H}}_{HV}$ replica space Hamiltonian. The reference energy for this point is $E/N = -3.169946 \pm 0.00004$. The parameters are $h_{x}=3$, $L=10$ and $\chi=2$}
\begin{tabular}{l l l l l l}
$\bar{\mathcal{O}}_{ZZ,HV}$ & $\bar{\mathcal{O}}_{ZZ,DD'}$  & $\bar{\mathcal{O}}_{X}$ & $E/N$ & $N\text{var}/E^{2}$ & $\delta E/E$ \\
\hline
\hline
 $\times$ & $\times$ & & -3.1682 & 0.0013  & 5.5 $10^{-4}$ \\
\hline
 $\times$ & $\times$ & $\times$ & -3.1682 & 0.0013 & 5.5 $10^{-4}$ \\
\hline
 $\times$ & &$\times$ & -3.163 & 0.005 & 2.0 $10^{-3}$ \\
\hline
&  $\times$ & $\times$ & -3.1682 & 0.0013 & 5.5 $10^{-4}$ \\
\hline
 &  &  & -3.163 & 0.005 & 2.0 $10^{-3}$ \\
\end{tabular}
\end{table}

\begin{figure}[h]
    \centering
    \includegraphics[width = 0.7\linewidth]{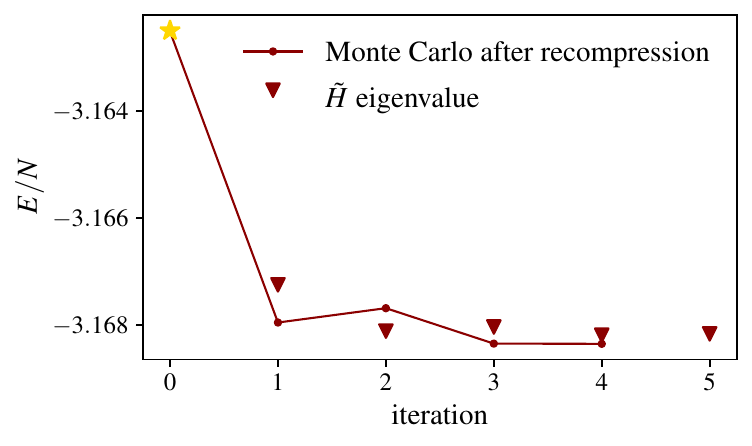 }
    \caption{Dual Monte Carlo optimization steps starting from the solution of $\bar{\mathcal{H}}_{HV}$ (star). At each step, Eq.~\eqref{eq:gen_eig2} is solved. The eigenvalues of the equation (arrows) are compared with the actual energies of the state after the compression (points).\label{fig:dmc-opti} $L=10$, $\chi=2$, $h_x=3$, and $10^{4}$ samples. 
    }
\end{figure}

\begin{figure}[h]
    \centering
    \includegraphics[width = 0.7\linewidth]{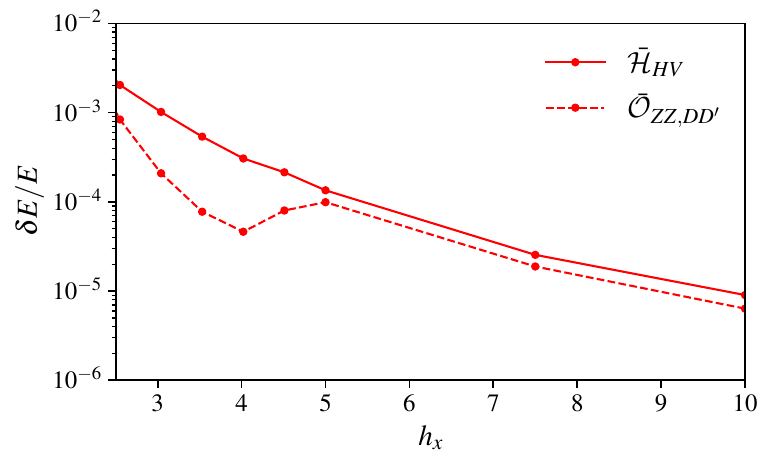}
    \caption{Error of $L=10$, $\chi=2$ RTT versus $h_x$ in the paramagnetic region of the phase diagram. The parameters and optimization procedure are the same as in Table \ref{tab:different-operators}: (solid line) replica space optimization using $\bar{\mathcal{H}}_{HV}$, followed by (dashed line) physical space optimization using $\bar{\mathcal{O}}_{ZZ,DD'}$. The error reported is an estimation using the V-score, $\delta E/E \approx 0.2 N \text{var}[H]/E^2$, see Ref.~\cite{srdinsek_hybrid_2026}.
    \label{fig:phase-diagram}}
\end{figure}

We now turn to physical space optimization using the dual Monte Carlo technique. We start from the RTT $\psi$ obtained using replica space optimization with the $\bar{\mathcal{H}}_{HV}$ replica space Hamiltonian Eq.\eqref{eq:H_HV} and seek the best linear combination Eq.\eqref{eq:lin_comb} using certain operators. Based on the results of the replica space optimization, we expect the important operators to be those that involve the diagonal part of the RTT but we will work in an agnostic way to see how the method works in practice. We will use $4$ operators: identity, $\bar{\mathcal{O}}_{X}$, $\bar{\mathcal{O}}_{ZZ,HV} = \bar{\mathcal{O}}_{ZZ,H\to H} + \bar{\mathcal{O}}_{ZZ,V\to V}$ and $\bar{\mathcal{O}}_{ZZ,DD'} = \bar{\mathcal{O}}_{ZZ,D\to D} + \bar{\mathcal{O}}_{ZZ,D'\to D'}$. The results are shown in Table \ref{tab:different-operators} for various choices of the operators included in the calculation. We find an improvement (of the same order as in the replica space optimization when we considered the $\bar{\mathcal{H}}_{HVDD'}$ replica space Hamiltonian) whenever the diagonal operator $\bar{\mathcal{O}}_{ZZ,DD'}$ was included. The main difference with replica space optimization is that in the physical space optimization, we can only improve the energy since we are building, by construction, the best linear combination of the different orbitals. We checked explicitly that the accuracy was not affected by the final compression used to put the linear combination back into an RTT form. Table \ref{tab:different-operators} also includes the rescaled variance of energy, a quantity known as the V-score \cite{wu_vscore_2024}. These results confirm what was surmized above: the information on the horizontal/vertical paths and transverse field was already included in the replica state optimisation so only the diagonal operator brings an improvement.

The above approach could be extended in various ways by considering more operators. For instance using powers of $\bar{\mathcal{H}}$, we can build a Lanczos-like method. We could also add operators that favor a given phase (say a staggering operator for the antiferromagnetic phase). We leave the associated explorations to further studies.

Fig.~\ref{fig:dmc-opti} shows different steps of the physical space optimization process. We observe that the convergence happens within a handful of iterations. Also, the eigenvalues of the generalized eigenproblem almost coincide with the result of the Dual Monte Carlo calculation which indicates that the compression performed after solving the linear system does not significantly alter the state.

Fig. \ref{fig:phase-diagram} shows the error obtained using the same optimization as in Table \ref{tab:different-operators} for a span of the $h_x$ values in the paramagnetic region of the phase diagram. We find that the error is consistently below $\delta E/E\approx 10^{-3}$. We did not study the antiferromagnetic side of the phase diagram which, we surmise, would require the introduction of operators that explicitly break translational symmetry since we expect a boundary phase transition before the bulk one \cite{kalinowski_bulk_2022}. An interesting feature of Fig. \ref{fig:phase-diagram} is a relatively large gain in precision obtained in the presence of the diagonal operator $\bar{\mathcal{O}}_{ZZ,DD'}$ around $h_x=4$. We attribute this feature to the increase of the antiferromagnetic correlation length as one decreases $h_x$: at some point it becomes large enough for the presence of the diagonal correlations to play a significant role. Lowering $h_x$ further, one would need to add more orderings and/or increase the bond dimension to keep up with the increasingly long-range correlations that build in the system.

\subsection{A study of RTT expressivity}
\begin{figure}[h]
    \centering
    \includegraphics[width = 0.7\linewidth]{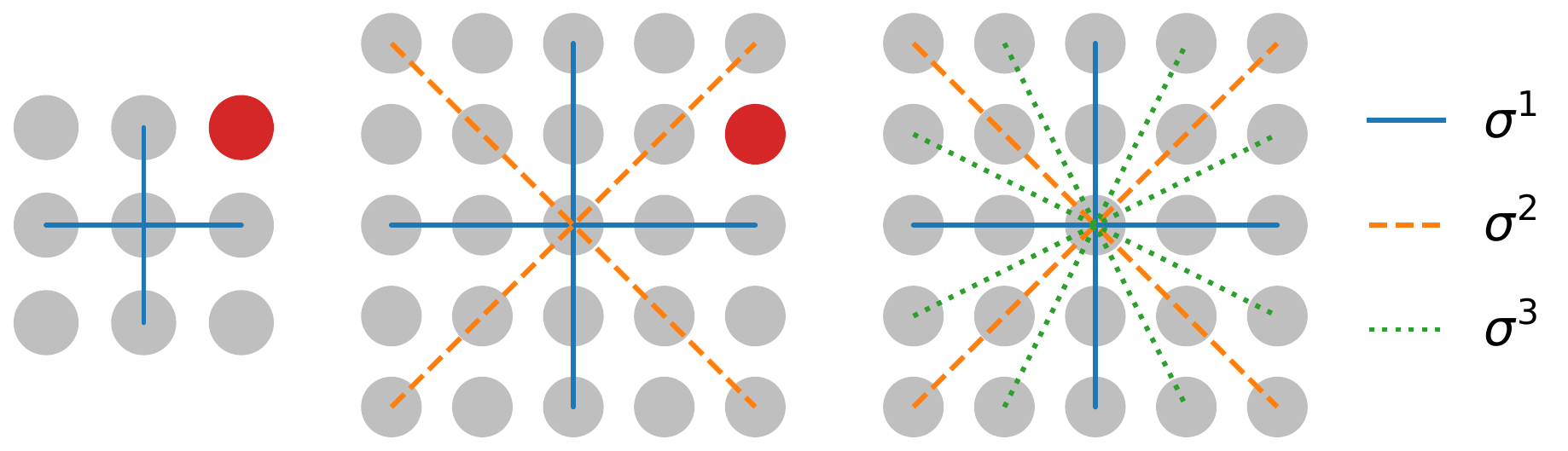}
    \caption{ Illustration of the sites that can be reached using the diffrent paths. Left, $\sigma^1 = \sigma^{HV}$; Middle, $\sigma^1 \oplus \sigma^2  = \sigma^{HVDD'}$; Right, $\sigma^1 \oplus \sigma^2 \oplus \sigma^3$ is new to this section. The sites shown in red are the closest sites to the central one that \emph{cannot} be reached from the paths.
    \label{fig:large-d-path}
    }
\end{figure}

The most striking aspect of the previous results is that a very precise energy $\delta E/E\approx 10^{-3}$ could be obtained with a very low rank $\chi=2$. A natural question is whether this error can be further decreased  and by which mechanism (rank increase or longer paths). In this section we generalize our RTT ansatz, increasing the two parameters that control its expressivity, $\Omega$ and $\chi$. There are two sides to the question of reaching higher accuracy: the expressivity of the RTT ansatz itself (with more complex paths and higher rank $\chi$) and the optimization process (choice of operators, virtual versus physical optimization etc). Here we study only the first aspect and use plain VMC with gradient descent on the RTT cores for the optimization. Studying the second aspect would require a study of the importance of different operators which we leave for subsequent publications on more challenging models.

Let us start by extending the scope of the studied paths. For our 2D model, a physical site is a point $(x,y)$ in the 2D grid so that we write $\sigma(\omega) = (x(\omega),y(\omega)$. For most values of $\omega$, the horizontal path correspond to $\sigma^H(\omega+1) = \sigma^H(\omega) + (1,0)$ (we ommit the transition from one line to the next on the boundaries for clarity, see Eq.\ref{eq:orderings}). Similarly, the vertical path corresponds to $\sigma^V(\omega+1) = \sigma^V(\omega) + (0,1)$; the diagonal ones to
$\sigma^D(\omega+1) = \sigma^D(\omega) + (1,1)$ etc.

We label these paths according to the Manhattan displacement distance $d$ on the lattice. $\sigma^1$ contains all distance $1$ displacements, i.e. $(0,1)$ and $(1,0)$. With our previous notations $\sigma^1 =\sigma^H \oplus \sigma^V = \sigma^{HV}$. Path $\sigma^1$ also potentially permit for correlations that connect a physical site to \emph{some} of its $d=2$ next nearest neigbours, i.e. $(0,\pm 2)$ and $(\pm 2,0)$ if the TT has a high enough rank. However, it cannot easily describe correlations with sites in the diagonal with displacement $(\pm 1,\pm 1)$ as these correlations correspond o virtual sites that are $\sim N$ away. $\sigma^2$ contains all distance $2$ displacements that are not in $\sigma^1$, i.e. $\sigma^2 (\omega+1) = \sigma^2(\omega) + (\pm 1,\pm 1)$. With our previous notations $\sigma^2 =\sigma^D \oplus \sigma^{D'}$ so $\sigma^1 \oplus \sigma^2$ is the most complex path of the previous section. We continue with
$\sigma^3$ that contains all the distance $3$ displacements that connot be obtained by following path $\sigma^1$ or $\sigma^2$: 
$\sigma^3 (\omega+1) = \sigma^3(\omega) +(\pm1,\pm2)$ and $(\pm2,\pm1)$. The full path up to distance $3$ is  $\sigma^1 \oplus \sigma^2 \oplus \sigma^3$. We also include $\sigma^4$ and $\sigma^5$ with similar definitions.

The intuition for this construction is illustrated in Fig.~\ref{fig:large-d-path} which shows which sites can be reached from the central one using the different paths. The red site is the closest one that \emph{cannot} be reached directly from the center. As we go from left ($\sigma^1$) to right ($\sigma^1 \oplus \sigma^2 \oplus \sigma^3$) we observe that this red site moves away from the central one so that all local correlations can easily be described. We therefore anticipate that improving the accuracy requires increasing both $d$ and $\chi$: increasing $d$ moves the unreachable (red) sites away while increasing $\chi$ is needed for the information to travel on longer distances in the path.

\begin{figure}[h]
    \centering
    \includegraphics[width = 1.0\linewidth]{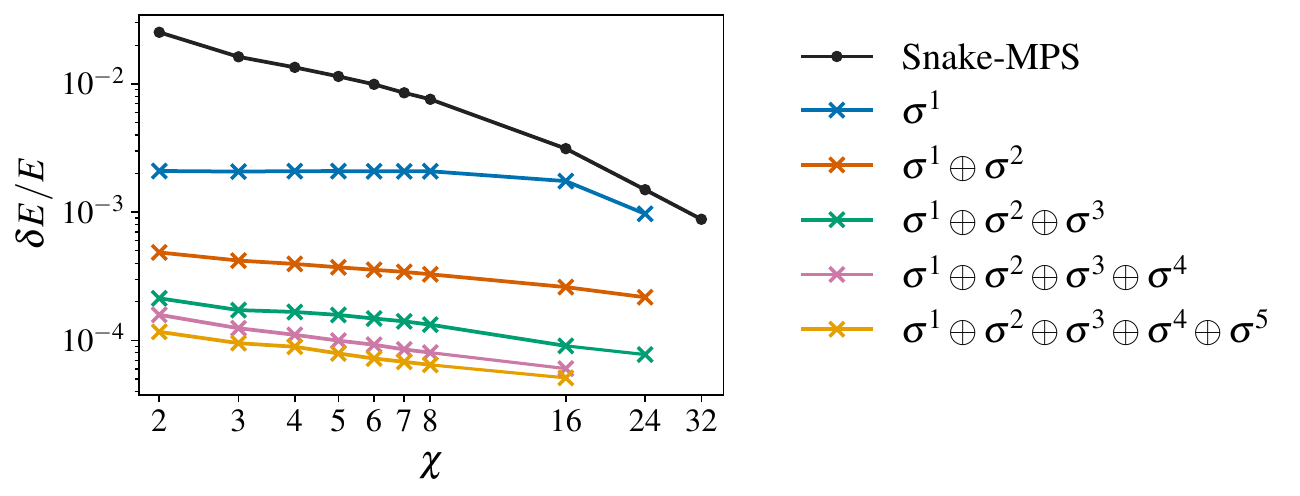}
    \caption{Relative energy error of $L=10$ at the critical point $h_x=3$ versus bond dimension $\chi$ with RTT for the cumulative paths $\sigma^1\oplus\cdots\oplus\sigma^d$. The crosses indicate different distances from $d=1$ (top) to $d=5$ (bottom). The black circles corresponds to a reference DMRG calculation that uses a "snake-MPS". 
    \label{fig:path-vs-rank}}
\end{figure}

To check the above intuition, we have optimized various RTT from $d=1$ to $d=5$
for the critical point $h_x=3$. We have used the following procedure: For each distance $d$, we construct a $\chi=2$ RTT($d$) by extending the optimized $\chi=2$ RTT($d-1$) with a rank one constant function for the new part of the path. Then, we optimize the 
obtained RTT (using standard stochastic gradient descent). Once we reach a plateau, we increase the rank and continue the optimization.
Interestingly, the largest fraction of the optimization steps is spent optimizing $\sigma^1$, the subsequent optimizations only require around 10\% as many steps as the first one.

The results are shown in Fig. \ref{fig:path-vs-rank}, where we report the relative energy error $\delta E/E$ versus $\chi$ for different values of $d$. We make several observations. First and most importantly, increasing $d$ and/or $\chi$ provides a systematic way to decrease the error, from $2.\ 10^{-3}$ down to
$5.\ 10^{-5}$, by almost two orders of magnitude. Second, increasing $\chi$ is ineffective for $\sigma^1$ unless $\chi$ becomes "so high" that even a simple MPS reaches the same accuracy (more on the significance of "so high" in the next section, it is expected to scale exponentially with system size). In contrast, increasing $d$ by one unit provides a large drop in the error. These two observations are consistent with our hypothesis that the error is dominated by the difficulty to correlate physical sites that are situated in diagonal (the red site in the left of Fig.~\ref{fig:large-d-path}). Third, we observe that the $d=2$ curve has a much stronger slope than $d=1$. Again, this is consistent with the fact that the next correlation that need to be captured are the next nearest neighbour along the vertical and horizontal directions, that are already reachable within $\sigma^1$ provided $\chi$ is big enough to convey the information. Overall these findings are qualitatively consistent with the intuition behind the schematic of 
Fig.~\ref{fig:large-d-path}.

\subsection{Algorithm complexity and discussion of other methods}
\label{sec:other}

As stated earlier, the 2D transverse field Ising model is one of the simplest genuine many-body problem; it can be solved with a variety of techniques including neural quantum states  (RBM \cite{carleo_solving_2017}, Transformer \cite{sprague2024transformer}), Projector Monte Carlo \cite{inglis2013}, Green function Monte Carlo  \cite{srdinsek_hybrid_2026}, stochastic series expansion \cite{koziol2021}, Path Integral Monte Carlo \cite{huang2020worm}, Tree tensor networks
\cite{Tagliacozzo2009} and even plain DMRG for systems not too large \cite{ganahl2023dmrgtpu}. In a recent publication, the authors have solved a similar model for a system as large as 100,000,000 spins using a taylored ansatz and variational Monte Carlo \cite{gouraud2026}. Our goal at this stage is therefore not to claim any superiority of the RTT (this, we hope, will come in subsequent publications) but more modestly to establish the technique as a new path to ground state exploration.

The number of free parameters in the ansatz is equal to $\sim\Omega \chi^2$
which is also the computational complexity of evaluating the RTT on a given configuration. From a given configuration, $\mathbf{s}$, there are $\sim N$ configurations $\mathbf{s'}$ for which the Hamiltonian $H_\mathbf{ss'}$ matrix element is non-zero. Using $N_w$ Monte Carlo samples requiring $O(N)$ Monte Carlo steps for thermalization, the overall cost of the stochastic part of the algorithm is $\sim N_w N \Omega \chi^2 \sim N_w K N^2\chi^2$ for a single matrix elements and $\sim N_w N \Omega \chi^2 M^2$ for the full eigenvalue problem. For the parameters used in this study, with a $N_w=10^4$, that translates into $10^{11}$ (embarrasingly parallel) operations per steps, a few seconds on a modern multicores CPU. The algebraic part of the algorithms consists of adding $\sim N$ RTT together, a process whose complexity scales as $\Omega \chi^3$. Overall replica optimisation scales as $\Omega N \chi^3$ per steps, i.e. faster than the Monte Carlo part for $\chi\ll N_w$. However, it typically requires many more step which compensates for the difference in complexity.
In our numerical experiments on a computer with 64 cores, for the $L=20$, $h_\text{x}=3$ data point, the Replica optimization converges in $15-20$ mn of CPU time while the subsequent Physical optimization only takes $1.5$ mn. For comparaison, obtaining the same accuracy with stochastic gradient descent took roughly $3$ hours.

With regards to VMC schemes based on (stochastic) gradient descent, the putative advantage of RTT is twofold. First, typical gradient descent traces require a very large number of Monte Carlo steps. This is highly dependent on the problem, ansatz and optimization methods but tens of thousand steps are typical in machine learning.  This is to be contrasted with the rapid convergence of replica optimisation and the handful of steps required by the physical optimisation. Second, so far NQS struggle to learn the sign of the wavefunction in models where a sign problem is present (our ultimate target). A common strategy is to use a smart NQS architecture that already captures this sign to a degree (e.g. using backflow corrections or similar ideas) but recent work indicates that this may come with a bias \cite{chen2026}. Here the hope, that remains to be demonstrated, is that the new techniques associated with the RTT, together with its own peculiar expressivity, will allow one to make progress in learning the sign of the wavefunction.

With regards to DMRG, it is expected that the rank needed to address a 2D system with a simple MPS scales as $\chi\sim \exp(L)$ due to the area law. For a $20\times 20$ lattice, it is still feasible to brute force the calculation. 
Ref.~\cite{ganahl2023dmrgtpu} actually did it using TPUs (i.e. large computational resources) and very large rank up to $\chi=65, 536$ exactly for our reference point $h_x=3$. They report a variational energy $E/N=-3.18197(2)$ consistent with ours within our stated error levels. We have repeated these DMRG calculations for smaller bond dimensions. The important point is that, with RTT,  we report an accuracy $\delta E/E\sim 10^{-3}$ that is essentially \emph{independent} of $L$ at fixed $\chi=2$. In contrast, in order to reach the same accuracy $\delta E/E\sim 10^{-3}$ with MPS, one requires $\chi=32$ ($L=10$) and $\chi=256$ ($L=20$) i.e. a factor $64$ increase in the number of parameters of the ansatz. RTT reaches the same accuracy with respectively $1.5\%$ ($L=10$) and $0.02\%$ ($L=20$) of the parameter count. These data validate the claim that RTT carries volume law entanglement.

With regards to PEPS and in particular to the recent trend to use Monte Carlo and gradient descent to reduce the computational burden associated with them 
\cite{wang2011,
liu_accurate_2021,
liu_tensor_2024,
naumann_rizzi_introduction_2024} 
(one only needs to contract the wave function, not its square, to sample), the RTT ansatz is much lighter computationally. Its evaluation is barely ($K\times$) more complicated than that of a MPS. At this stage, however, we cannot compare the expressivity of the two models nor their trainability.

\section{Conclusions}
\label{sec:Conclusions}

In this work we have introduced a variational ansatz, the Replica Tensor Train, that naturally extends a class of previously discussed powerful ans\"atze \cite{schuch_strings_2008,sfondrini_simulating_2010,glasser_neural-network_2018,srdinsek_hybrid_2026}. Our chief result is that RTT has an internal structure that provides the possibility to build the ground-state of a Hamiltonian directly without resorting to gradient descent. We have shown a proof of principle that the method works for the 2D Ising model with transverse field, obtaining a very good accuracy while using a minimal bond dimension $\chi=2$.

This article opens more new questions than it answers and much more work is needed. The most urgent question is whether the set of new techniques will be efficient to solve more difficult models, notably models with sign problems. Indeed, while VMC is immune to the sign problem, we do not yet have a general approach to \emph{learn the sign}. The hope is that our \emph{global} operator based approach will not suffer from the same limitations as \emph{local} gradient descent. One could also envision combining the two approaches. While we focused on ground-state search, the dynamics, within the time-dependent variational principle framework should also be directly in scope of the operator approach (in either replica and/or physical space). On the technical ground, one needs to study how the quality of the ansatz can be controlled using e.g. higher ranks, more complex paths, more operators and possibly meta algorithms that choose the best combination dynamically. Last, variants of the RTT-based techniques compatible with quantum hardware could also be constructed.

\section*{Acknowledgements}
MS would like to thank Antoine Georges' group at Collège de France for hosting him during his stay in Paris.

\paragraph{Author contributions}
MS wrote the code and performed the simulations. MS, GG and XW contributed to the mathematical concepts, algorithms, data analysis and writing of the manuscript. XW supervised the project.

\paragraph{Funding information}
We acknowledge the Plan France 2030 ANR-22-PETQ-0007 “EPIQ”, the PEPR “EQUBITFLY”, the ANR “DADI” and the CEA-FZJ French-German project AIDAS for the funding.

\bibliography{Citations.bib}

\end{document}